\documentclass[prl,twocolumn,showpacs,superscriptaddress,floatfix]{revtex4-1}
\usepackage{graphicx,amsfonts,amssymb,amsmath,xspace,dsfont}
\usepackage[colorlinks=true,citecolor=green,linkcolor=red,urlcolor=blue]{hyperref}
\usepackage{color}
\definecolor{g}{rgb}{.1,0.4,.1} 
\definecolor{b}{rgb}{0,0.2,1}
\definecolor{rouge}{rgb}{0.82,0.,0.}
\definecolor{vert}{rgb}{0.,0.82,0.}
\definecolor{orange}{rgb}{1,0.5,0.}
\definecolor{bleu}{rgb}{0.,0.,0.82}
\definecolor{m}{rgb}{0.82,0.,0.82}
\definecolor{vert2}{rgb}{0.,0.5,0.}
\definecolor{rougeclair}{rgb}{1.0,0.7,0.7}

\begin{document}

\title{Persisting topological order via geometric frustration}

\author{Kai Phillip Schmidt}
\email{kai.schmidt@tu-dortmund.de}
\affiliation{Lehrstuhl f\"{u}r Theoretische Physik I, Otto-Hahn-Stra\ss e 4, TU Dortmund, 44221 Dortmund, Germany}

\begin{abstract}
We introduce a toric code model on the dice lattice which is exactly solvable and displays topological order at zero temperature. In the presence of a magnetic field, the flux dynamics is mapped to the highly frustrated transverse field Ising model on the kagome lattice. This correspondence suggests an intriguing disorder by disorder phenomenon in a topologically ordered system implying that the topological order is extremely robust due to the geometric frustration. Furthermore, a connection between fully frustrated transverse field Ising models and topologically ordered systems is demonstrated which opens an exciting physical playground due to the interplay of topological quantum order and geometric frustration.     
\end{abstract}


\maketitle

%
%
\emph{Introduction ---}
%
%
Topological quantum order, as introduced by Wen in the context of high-temperature superconductivity \cite{Wen89_1,Wen90_1}, 
has become a very active research topic in recent years, since it plays an important role for the 
 physics of the fractional quantum Hall effect, for frustrated magnetism, and in the field of quantum
 information due to the fascinating perspective to build a topological quantum computer which is protected 
from local decoherence \cite{Kitaev03,Ogburn99}. 

One of the standard models displaying all essential features of topological
 quantum order at zero temperature, e.g.~a ground-state degeneracy depending on the genus and elementary 
excitations with fractional statistics, is Kitaev's toric code \cite{Kitaev03} which is an exactly solvable two-dimensional 
quantum spin model. The toric code therefore represents a perfect starting point to study fundamental 
properties of topologically ordered quantum systems, e.g. several works have studied the robustness 
and the ascociated topological phase transitions of the toric code on the square lattice in the presence of a magnetic field 
\cite{Trebst07,Hamma08,Vidal09_1,Vidal09_2,Tupitsyn10,Dusuel11,Wu12}. 

Although the full phase diagram of the toric code in a magnetic field is very rich \cite{Dusuel11}, one finds generically 
 a quantum phase transition in the 3d Ising universality class between the topologically ordered 
phase and a conventional polarized phase where spins are aligned along the field direction. This
 behaviour is best understood for a single parallel field where the well-known duality between
 $\mathbb{Z}_2$ gauge theories and  {\it unfrustrated} transverse field Ising models (TFIMs) applies \cite{wegner71,fradkin79,kogut79,Trebst07,Hamma08}. Physically, the parallel magnetic field induces kinetic energy into the system and the quantum phase transition out of the topological phase corresponds to a condensation of elementary charge or flux excitations of the topological phase which live on the dual lattice \cite{fradkin78,kogut79}.

Hence, one might wonder whether also duality mappings between toric code models in a field and 
 {\it frustrated} Ising models exist. This would imply that the kinetics of the elementary excitations
 in the topological phase is strongly reduced due to the frustration and, as a consequence, 
the topological order is expected to be 
very robust. Furthermore, the very rich physics of fully frustrated TFIMs 
should be also present in systems displaying topological quantum order, e.g.~topological phase transitions 
in different universality classes are expected \cite{moessner2000,moessner2001}. 
%
\begin{figure}[t]
\centering
\includegraphics[width=\columnwidth]{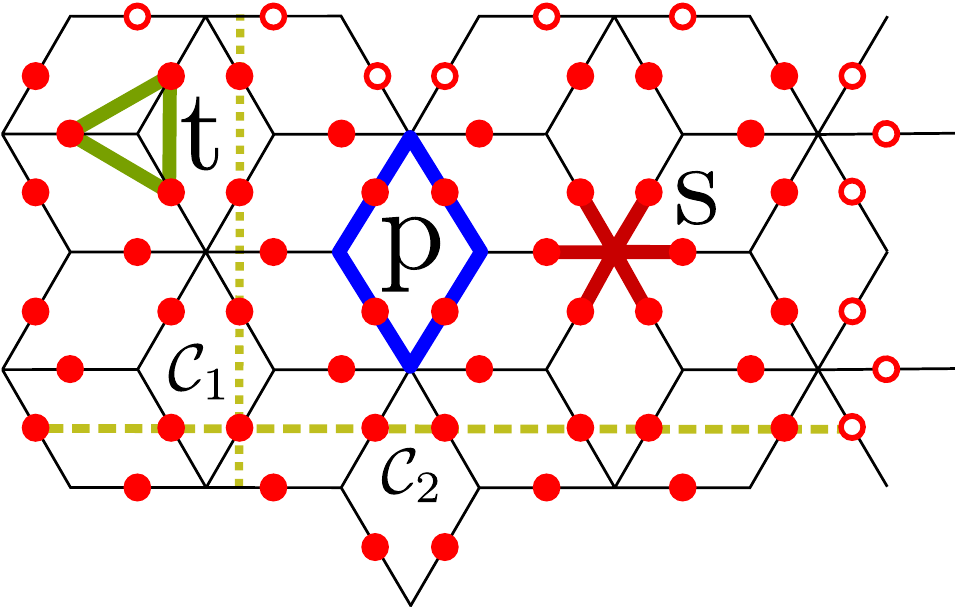}
\caption{(color online). Illustration of the toric code model on the dice lattice (thin black lines) with periodic boundary conditions on a torus. Spin 1/2 degrees of freedom are located on the links of the lattice and are denoted by big red dots. The links with open red circles on the upper (right) end of the cluster have to be identified with corresponding links on the lower (left) end. Operators $X^{t}$, $X^{s}$, and $Z_p$ are defined on triangles $t$, stars $s$, and plaquettes $p$. The non-contractible loop operators $\mathcal{Z}_\mu$ with $\mu\in\{1,2\}$ are defined on contours $\mathcal{C}_1$ (dotted line) and $\mathcal{C}_2$ (dashed line).}
\label{fig:1}
\end{figure}
%

Most extremely, the TFIM on the kagome lattice realizes a {\it disorder by disorder} scenario \cite{moessner2000,moessner2001,powalski2013}, 
i.e.~the ground state is quantum disordered for any value of the field. Assuming a mapping of the kagome TFIM 
to a toric code in a magnetic field, implies an ultimatively robust topological quantum order for {\it any} value
 of the field due to the strong geometric frustration. In other words, quantum fluctuations induced by the toric code on the 
extensively-many classical ground states select a topologically ordered spin liquid ground state, a scenario first suggested by Anderson and Fazekas \cite{anderson73,anderson74}. 

In this letter, we present such a fascinating disorder by disorder scenario for the toric code on the dice lattice. Furthermore, we demonstrate a general connection between perturbed toric codes and fully frustrated TFIMs.

%
\emph{Model ---}
%
%
The Hamiltonian of the toric code on the dice lattice is given by
%
\begin{equation}
 \label{eq:ham}
 H = -J^{t}\sum_{t} X^{t} -J^{s}\sum_{s} X^{s} -J^{p}\sum_{p} Z^{p},
\end{equation} 
%
%
where $t$ refers to triangles, $s$ to stars, and $p$ to plaquettes as displayed in Fig.~\ref{fig:1}. The corresponding operators are defined by $X^{t}=\prod_{i \in t} \sigma_i^x$, $X^{s}=\prod_{i \in s} \sigma_i^x$, and \mbox{$Z^{p} =\prod_{i \in p} \sigma_i^z$} where the $\sigma_i^\alpha$'s with $\alpha\in x,y,z$ are the usual Pauli matrices. Plaquettes always share an even number of sites with triangles or stars and one finds $[Z^{p},X^{t}]=[Z^{p},X^{s}]=[X^{t},X^{s}]=0$ for any plaquette $p$, triangle $t$, and star $s$. The eigenvalues $z^{p}=\pm 1$, $x^{t}=\pm 1$, and $x^{s}=\pm 1$ of these operators are therefore conserved quantities which allows an exact solution of Eq.~(\ref{eq:ham}).  

%
\emph{Ground states ---}
%
%
In the following we assume $J^{\alpha}>0$ with $\alpha\in t,s,p$. As a consequence, ground states correspond to states having all eigenvalues $z^{p}=x^{t}=x^{s}=+1$. The number of ground states depends on the genus and the system is topologically ordered. To be explicit, the ground state is unique for an open plane
%
\begin{equation}
 \label{eq:gs_open}
 |0\rangle =  \frac{1}{\mathcal{N}} \prod_t \frac{\mathds{1}+X^{t}}{2} \prod_s \frac{\mathds{1}+X^{s}}{2} \prod_p \frac{\mathds{1}+Z^{p}}{2} |\Rightarrow\rangle ,
\end{equation} 
%
%
where $\mathcal{N}$ is a normalization constant and $|\Rightarrow\rangle$ corresponds to a fully polarized state where all spins point in $x$-direction. In contrast, on the torus one has two conserved non-contractible loop operators $\mathcal{Z}_\mu=\prod_{i\in\mathcal{C}_\mu}\sigma_i^z$ with $\mu\in\{1,2\}$ defined on contours $\mathcal{C}_\mu$ (see Fig.~\ref{fig:1}) and one obtains the four ground states 
%
\begin{equation}
 \label{eq:gs_torus}
|z_1,z_2\rangle = \left( \frac{\mathds{1}+z_1\mathcal{Z}_1}{2}\right)\left( \frac{\mathds{1}+z_2\mathcal{Z}_2}{2} \right) | 0\rangle ,
\end{equation} 
%
%
where $z_\mu\in\{\pm 1\}$ denotes the eigenvalues of the operators $\mathcal{Z}_\mu$. As for the conventional toric code \cite{Kitaev03}, one finds a ground-state degeneracy $4^{g}$ for a system with genus $g$.

%
\emph{Excitations ---}
%
%
Elementary excitations are states where one eigenvalue out of $x^{t}$, $x^{s}$, or $z^{p}$ is $-1$. In contrast to the conventional toric code on the square lattice, there are two type of charges, one on triangles ($x^{t}=-1$) and one on stars ($x^{s}=-1$). Additionally, there are fluxes on plaquettes having $z^{p}=-1$. All excitations are static and non-interacting because they are protected by conservation laws. Charges and fluxes have a mutual semionic statistics, i.e. charges and fluxes among themselves are hardcore bosons but one gets a non-trivial factor $-1$ when braiding charges around fluxes (or vice versa).

%
\emph{Toric code in an $x$-field ---}
%
%
Let us add a magnetic field $H_x=-h_x\sum_i\sigma_i^x$ to the toric code. Triangle and star operators still commute with the full Hamiltonian and their eigenvalues remain conserved quantities. The Hilbert space therefore separates into sectors where each block corresponds to a fixed configuration of charges. 

\begin{figure}[t]
\centering
\includegraphics[width=0.75\columnwidth]{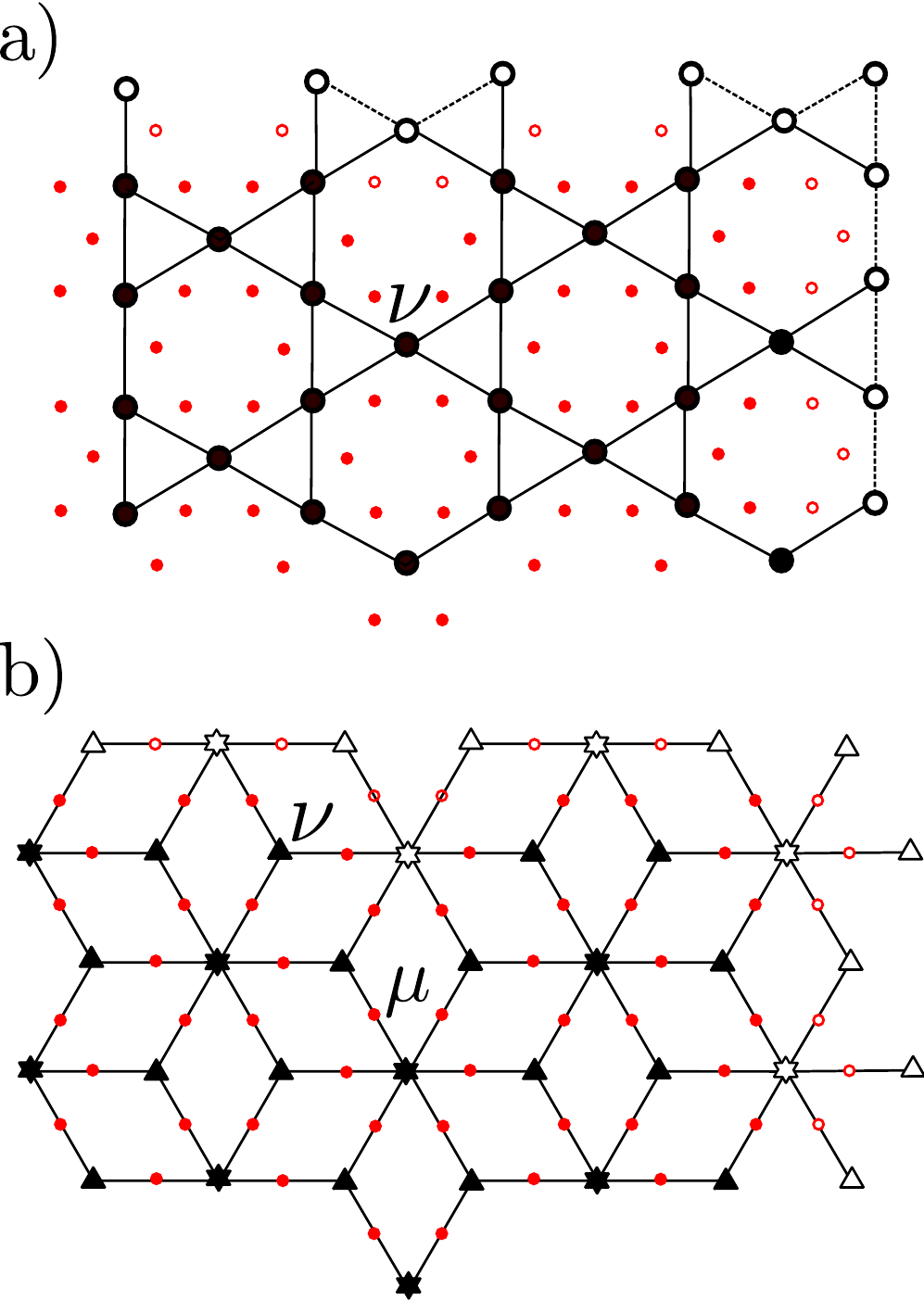}
\caption{(color online). Illustration of the (a) dual kagome lattice formed by the centers of plaquettes $p$ (black circles $\nu$) and of the (b) dual dice lattice build by the centers of triangles and stars (black triangles $\nu$ and stars $\mu$). The small red circles denote in both figures the location of the link variables on the original dice lattice displayed in Fig.~\ref{fig:1}.}
\label{fig:2}
\end{figure}

The topologically ordered ground state of the bare toric code ($h_x=0$) is in the charge-free sector having \mbox{$x^{t}=x^{s}=+1$} on all triangles and stars. The physics in this sector at finite fields depends only on the flux excitations. To be specific, the action of $\sigma_i^x$ on any eigenstate of the toric code results in flipping the eigenvalues $z^p$ of the two plaquette operators $Z^p$ which are attached to site $i$. As a consequence, the fluxes aquire a finite dispersion due to kinetic energy induced by the magnetic field $h_x$. Introducing pseudo-spin 1/2 operators $\tau^\alpha_\nu$ with $\alpha=x,y,z$ on the sites $\nu$ of the dual kagome lattice of plaquettes $p$ (see Fig.~\ref{fig:2}a), one obtains the kagome TFIM which is given by    
%
%
%
%
\begin{equation}
 \label{eq:ham_TFIM_Kagome}
 H = -J^{p}\sum_{\nu} \tau_\nu^z -h_x \sum_{\langle \nu,\nu^\prime\rangle} \tau_\nu^x \tau_{\nu^\prime}^x ,
\end{equation} 
%
%
where the second sum runs over all bonds of the kagome lattice. The presence or absence of a flux on plaquette $p$ is encoded in a pseudo-spin pointing in $+z$ or $-z$ direction. The topological phase is mapped to the polarized phase in the dual language. Consequently, the energetic properties of fluxes in the topological phase are identical to spin-flip excitations in the kagome TFIM.

The physics depends strongly on the sign of $h_x$, since the Ising coupling in Eq.~\ref{eq:ham_TFIM_Kagome} is ferromagnetic for $h_x>0$ and antiferromagnetic for $h_x<0$. The ferromagnetic case is rather conventional. Here, the model is unfrustrated, and one expects a second-order phase transition in the 3d Ising universality class between the topologically ordered phase and the polarized phase as for the conventional toric code on the square lattice. The low-energy physics is thus taking place in the flux-free sector for the whole ferromagnetic parameter axis, since also the polarized ground state is part of this sector. We have calculated the one-flux gap with momentum $\vec{k}=(0,0)$ up to order 13 in $h_x/J$ \cite{Supp_Mat}. Using dlog Pad\'{e} extrapolation \cite{Extrapol}, the quantum critical point is located at a ratio  $h_x/J=0.3390(5)$ with an exponent $\nu\approx 0.644$ fully consistent with a 3d Ising transition ($\nu_{\rm 3d\,Ising}\approx 0.630)$ \cite{Bloete95}. 

The antiferromagnetic case $h_x<0$ is fundamentally different due to the strong frustration. Translating the properties of the kagome TFIM to the perturbed toric code \cite{powalski2013}, the flux excitation remains gapped for any value of the field. The topological phase is therefore robust on the whole antiferromagnetic parameter axis in the charge-free sector. This robustness is a consequence of the frustrated kinetics of flux excitations resulting in a strongly reduced bandwidth for fluxes. In fact, the lowest flux band is completely flat up to order 7 in perturbation theory \cite{powalski2013}. 

It is also enlightening to interpret the physics from the limit $J^{p}\rightarrow 0$ which corresponds to the pure Ising model on the kagome lattice in the dual language. Here, the kagome TFIM realizes an intriguing disorder by disorder scenario \cite{powalski2013}. One has infinitely many classical ground states and an infinitesimally small $J^p$ selects a quantum disordered polarized phase as unique ground state. For the toric code in the presence of an antiferromagnetic $x$-field, one also has extensively many classical ground states for $J^{p}=0$ due to the charge-free constraint \mbox{$x^{t}=+1$} on all triangles. These states have exactly one spin on each triangle pointing in ($+x$)-direction which corresponds to an energy per site $e_0^{\rm cf}=(2h_x-2J^{t}-J^{s})/6$. The duality to the kagome TFIM then implies that the quantum fluctuations induced by the plaquette operators $Z^{p}$ directly select a quantum disordered topologically ordered ground state (a $\mathbb{Z}_2$ spin liquid) which persists for any value of $J^p$ up to the limit of the bare toric code.

Next we discuss under which conditions the low-energy physics of the toric code in an $x$-field is indeed contained in the charge-free sector. This question can be answered in the limit $J^{p}=0$ where all eigen states are classical states. A natural competitor to the charge-free states is then the fully-polarized state $|\Leftarrow\rangle$ where all spins point in ($-x$)-direction, since this state minimizes the field term for $h_x<0$. Its energy per site is given by $e_0^{\rm \Leftarrow}=(6h_x+2J^{t}-J^{s})/6$ because one has $x^{t}=-1$ and $x^{s}=+1$ for all triangles and stars. The classical charge-free states therefore have a lower energy than the polarized state $|\Leftarrow\rangle$ whenever $|h_x|<J^{t}$ for $J^{p}=0$. Consequently, it is always possible to enlarge the topological phase by increasing the ratio $J^{t}/|h_x|$.

%
\emph{Toric code in a $z$-field ---}
%
%
Now we consider the effect of a magnetic field $H_z=-h_z\sum_i\sigma_i^z$ added to the toric code. In this case the plaquette eigenvalues $z^p$ are still conserved quantities and each flux sector can be studied separately. The remaining degrees of freedom are then the charge excitations on triangles and stars which live on a dual dice lattice (see Fig.~\ref{fig:2}b).

The low-energy physics takes place in the flux-free sector as for the conventional toric code. Defining pseudo-spin 1/2 operators $\tau^\alpha_\zeta$ with \mbox{$\alpha=x,y,z$} on the sites $\zeta$ of the dual dice lattice of triangles $t$ and stars $s$, one obtains a TFIM on the dice lattice 
%
\begin{equation}
 \label{eq:ham_TFIM_dice}
 H = -J^{t}\sum_{\nu} \tau_\nu^z -J^{s}\sum_{\mu} \tau_\mu^z-h_z \sum_{\langle \nu,\mu\rangle} \tau_\nu^x \tau_{\mu}^x ,
\end{equation} 
%
%
where we distinguish between sites $\nu$ ($\mu$) of triangles (stars) as illustrated in Fig.~\ref{fig:2}b. For $J^{t}=J^{s}\equiv J$, one has the isotropic TFIM on the dice lattice which is unfrustrated and one expects a quantum phase transition in the 3d Ising universality class. This is confirmed by analysing the high-order series expansion of the one-charge gap with momentum $\vec{k}=(0,0)$ which we have determined up to order $8$ in $h_z/J$  \cite{Supp_Mat}. Using dlog Pad\'{e} extrapolation \cite{Extrapol}, the quantum critical point is located at $h_z/J~\approx 0.3095(2)$ with an exponent $\nu\approx 0.649$.  Since the dice lattice is bipartite, one gets the same kind of quantum critical point for negative fields at $h_z/J~\approx -0.3095(2)$. Similarly, one expects the same universal behaviour for anisotropic ratios $J^{t}\neq J^{s}$. 

%
\emph{Dualities to fully frustrated TFIMs ---}
%
%
The intriguing duality between the toric code in an $x$-field on the dice lattice and the antiferromagnetic kagome TFIM can be generalized to fully frustrated TFIMs on other lattices. Fully frustrated Ising models obey the constraint $\prod_{\rm plaquette} (-J_{ij}/J)=-1$ on every elementary plaquette of the lattice with the nearest-neighbor exchange $|J_{ij}|=J$ between sites $i$ and $j$, e.g.~on the square lattice one can take three ferromagnetic and one antiferromagnetic coupling on each plaquette. As a consequence, there exist extensively-many ground states for the pure Ising model due to the geometric frustration. In contrast to the kagome TFIM, a transverse magnetic field gives rise to an order by disorder scenario on most lattices, i.e. quantum fluctuations select an ordered (symmetry-broken) ground state from the degenerate manifold which then breaks down by a quantum phase transition to a polarized phase for larger values of the field \cite{moessner2000,moessner2001}. Assuming dualities between fully frustrated TFIMs and toric code models in a field, implies the existence of interesting quantum phase transitions between topologically ordered quantum matter and conventionally ordered phases.

In the following we give the explicit construction of such a duality mapping between the conventional toric code in a field on the square lattice and the fully-frustrated TFIM on the square lattice. This construction can be generalized to other lattices in a straightforward manner. To be specific, we study
%
%
\begin{equation*}
 \label{eq:ham_tc}
 H = - J_s \sum_{s} A_s  - J_p \sum_{p} B_p  - h_x\sum_i \sigma_i^x,
\end{equation*} 
%
%
where $A_s=\prod_{i \in s} \sigma_i^x$ and $B_p =\prod_{i \in p} \sigma_i^z$, and we assume $h_x>0$ without loss of generality. Subscript $s$ ($p$) refers to sites (plaquettes) of a square lattice and $i$ runs over all bonds where spins are located (see Ref.~\onlinecite{Kitaev03} for details). Charge eigenvalues $a_s=\pm 1$ of operators $A_s$ are therefore conserved quantities and the Hilbert space separates into different sectors of fixed charged configurations.  

Usually, positive couplings \mbox{$J_s=J_p=J>0$} are considered. In this case the low-energy physics takes place in the charge-free sector with $a_s=+1$ for all $s$, and the well-known duality between the $\mathbb{Z}_2$ gauge theory and the {\it unfrustrated} TFIM on the dual square lattice applies \cite{wegner71,fradkin79,kogut79,Trebst07,Hamma08}. As a consequence, a quantum phase transition in the 3d Ising universality class between the topologically ordered phase and the polarized phase is detected for $h_x/J\approx 0.328$ \cite{Bloete2002}. 

The physics is strongly different for the choice \mbox{$-J_s=J_p\equiv J$} with $J>0$. The topologically ordered ground state in the absence of the magnetic field $h_x=0$ is then in the {\it charge-full} sector $a_s=-1$ for all sites $s$ ($b_p=+1$ for all plaquette operators $B_p$). A finite magnetic field gives then rise to mobile fluxes which hop in the background of static charges. Performing the same kind of duality mapping as above by introducing pseudo-spin operators on the sites of the dual square lattice (of plaquettes), one finds that the charge-full sector is dual to the fully frustrated TFIM on the square lattice. Note that such dualities between odd Ising gauge theories and fully frustrated TFIM have been already discussed in the context of quantum dimer models \cite{moessner2001b}. 

The fully frustrated TFIM on the square lattice realizes a 3d XY quantum phase transition \cite{Blankschtein84,moessner2001,Xu2009,Wenzel2012} between the polarized phase and a columnar ordered phase at a critical ratio $h_x/J\approx 0.634$ \cite{Wenzel2012}. Consequently, the extension of the topological phase (corresponding to the polarized phase in the dual language) is again enlarged due to the frustration. Physically, the frustration in the TFIM is induced by the presence of the background charge in the charge-full sector of the topological phase.   

As for the perturbed toric code on the dice lattice, one has to investigate the role of other charge sectors for the low-energy physics. Indeed, the magnetic field term favors the fully polarized state $|\Rightarrow\rangle$ where all spins point in $(+x)$-direction. This state is in the {\it charge-free} sector and, as a consequence, one expects a first-order phase transition at a certain ratio $h_x/J$ when the charge-free sector contains the true ground state of the system. For $J_p=0$, the classical energies per site are given by $e_0^{\rm \Rightarrow}=(J-2h_x)/2$ and $e_0^{\rm  charge-full}=(-J-h_x)/2$ such that the charge-full sector contains the ground state for $J>h_x /2$. Thus, it is always possible to stabilize the charge-full sector including the 3d XY quantum phase transition by enlarging the ratio $J_s/h_x$. 

For the specific case \mbox{$-J_s=J_p\equiv J$} with $J>0$, we have calculated the ground-state energy per site $e_0^{\rm  charge-full}$ of the topological phase up to order 8 in $h_x/J$ which is equivalent to the ground-state energy per site of the polarized phase in the dual fully frustrated TFIM on the square lattice (up to the constant energy \mbox{$J_s/2=-J/2$} due to the presence of charges) \cite{Supp_Mat}. The energy $e_0^{\rm  charge-full}$ has to be compared to the ground-state energy per site of the charge-free sector. The latter sector is dual to the conventional unfrustrated TFIM on the square lattice (up to a constant energy offset \mbox{$-J_s/2=+J/2$} due to the absence of charges) and series expansions of $e_0$ for the ordered and disordered phase are available \cite{He1990,Oitmaa1991}. Altogether, by comparing both sectors, it can be clearly deduced that the low-energy physics takes place in the charge-full sector for $h_x/J$ well larger than the 3d XY quantum critical point at $\approx 0.634$ \cite{Wenzel2012}. Interestingly, the presence of the background charge, or equivalently the geometric frustration in the dual picture, leads to a different universality class compared to the charge-free (unfrustrated) case.  

%
%
\emph{Conclusions ---}
%
%
In this work we studied the impact of geometric frustration on topologically ordered quantum matter. 
It is found that the robustness of topological order is enhanced if the kinetics of anyonic excitations is
 reduced due to the geometric frustration. Furthermore, a plethora of fascinating phenomena including a disorder 
by disorder scenario for the toric code on the dice lattice are presented for topologically-order 
quantum phases and their breakdown which opens an interesting playground for future research.

KPS acknowledges ESF and EuroHorcs for funding through the EURYI as well as DFG. Furthermore, fruitful discussions 
with K. Coester, S. Dusuel, M. Kamfor, R. Moessner, M. Powalski, M.D. Schulz, G.S. Uhrig, J. Vidal, and C. Xu are acknowledged.


%


\onecolumngrid
\newpage      

\section*{Supplemental Material}
In the following, we give the series expansions in the polarized phase for the one-particle gap of the ferromagnetic TFIM on the kagome lattice, the one-particle gap of the isotropic TFIM on the dice lattice, as well as the ground-state energy per site of the fully-frustrated TFIM on the square lattice. Finally, we describe shortly the concept of dlogPad\'{e} extrapolation.

\subsection{One-particle gap of the ferromagnetic TFIM on the kagome lattice}

The kagome TFIM as introduced in the manuscript is given by    

%
%
%
\begin{equation}
 \label{eq:ham_TFIM_Kagome}
 H = -J^{p}\sum_{\nu} \tau_\nu^z -h_x \sum_{\langle \nu,\nu^\prime\rangle} \tau_\nu^x \tau_{\nu^\prime}^x ,
\end{equation} 
%
%
where $\tau^\alpha_\nu$ are pseudo-spin 1/2 operators with $\alpha=x,y,z$ on the sites $\nu$ of the kagome lattice. For a ferromagnetic coupling $h_x>0$, the one-particle gap $\Delta$ of the polarized phase has a momentum $\vec{k}=(0,0)$. Its series expansion is given by 

%
%
\begin{align}
	\Delta / J^p =& \ 2-4x-2x^2 -\frac{3}{2} x^3-\frac{9}{2} x^4 -\frac{261}{32} x^5
	 -\frac{2523}{128} x^6  -\frac{41891}{1024} x^7  -\frac{760415}{8192} x^8 -\frac{14668135}{65536} x^9  \nonumber\\&\
	-\frac{450811297}{786432} x^{10}-\frac{18337562173}{12582912} x^{11}-\frac{561320588059}{150994944} x^{12}-\frac{78314757058910141}{8246337208320} x^{13} \,, \nonumber\\
\end{align}
%
%
where $x=(h_x/ J^p)$.

\subsection{One-particle gap of the isotropic TFIM on the dice lattice}

The TFIM on the dice lattice as introduced in the manuscript is given by    
%
\begin{equation}
 \label{eq:ham_TFIM_dice}
 H = -J^{t}\sum_{\nu} \tau_\nu^z -J^{s}\sum_{\mu} \tau_\mu^z-h_z \sum_{\langle \nu,\mu\rangle} \tau_\nu^x \tau_{\mu}^x ,
\end{equation} 
%
%
where sites $\nu$ ($\mu$) corresponds to triangles (stars) on the dice lattice. For $J^{t}=J^{s}\equiv J$, one has the isotropic TFIM on the dice lattice. For postive $h_z>0$, the one-particle gap $\Delta$ of the polarized phase has a momentum $\vec{k}=(0,0)$ and its series expansion is given by 

%
%
\begin{align}
	\Delta / J =& \ 2 -3\sqrt{2}x -\frac{9}{4}x^2 -\frac{159\sqrt{2}}{64} x^3 -\frac{99}{16} x^4 -\frac{85461\sqrt{2}}{8192} x^5 -\frac{32067}{1024} x^6 -\frac{31283943\sqrt{2}}{524288} x^7 -\frac{6725967}{32768} x^8  \,, \nonumber\\
\end{align}
%
%
where $x=(h_z/ J)$.

\subsection{Ground-state energy per site of the fully-frustrated TFIM on the square lattice}

The fully-frustrated TFIM on the square lattice as introduced in the manuscript is given by
%
\begin{equation}
 \label{eq:ham_TFIM_square}
 H =  -J \sum_{\nu} \tau_\nu^z -h_x \sum_{\langle \nu,\nu^\prime\rangle} \tau_\nu^x \tau_{\nu^\prime}^x ,,
\end{equation} 
%
%
where $\nu$ denotes the sites of a square lattice. The series expansion of the ground-state energy per site  $e_0$ in the polarized phase is given by 

%
%
\begin{align}
	e_0 / J =& -1-\frac{1}{2} x^2+\frac{5}{32} x^4-\frac{31}{128} x^6+\frac{1531}{4096} x^8  \,, \nonumber\\
\end{align}
%
%
where $x=(h_x/ J)$.

\subsection{DlogPad\'{e} extrapolation}

We use dlogPad\'e extrapolation to analyze the series expansion of a one-particle gap $\Delta$ in a perturbative parameter $x$. Different extrapolants $\left[L,M\right]$ are constructed where $L$ denotes the order of the numerator and $M$ the order of the denominator. Explicitly, the dlogPad\'e extrapolation is based on the Pad\'{e} extrapolation of the logarithmic derivative of the one-particle gap $\Delta$
\begin{equation}
 \left[\frac{d}{dx}\ln \Delta \right]_{[L,M]}:=\frac{P_{L}}{Q_M}\quad ,
 \label{eq:dlog}
\end{equation}
where $P_{L}$ and $Q_M$ are polynomials of order $L$ and $M$. Due to the derivative of the numerator in Eq.~\ref{eq:dlog} one requires $L+M=m-1$ where $m$ denotes the maximum perturbative order which has been calculated. The $\left[L,M\right]$ dlogPad\'e extrapolant is then given by
\begin{equation}
 \left[L,M\right] :=\exp\left(\int_0^x \frac{P_{L}(x')}{Q_M(x')} dx'\right)\quad .
 \label{eq:dlog2}
\end{equation}
In the case of a physical pole at $x_0$ one is able to determine the dominant power-law behaviour $|x-x_0|^{z\nu}$ close to $x_0$. The exponent $z\nu$ is then given by the residuum of $P_L/Q_M$ at $x=x_0$
\begin{equation}
 z\nu =\frac{P_{L}(x)}{\frac{d}{dx}Q_M(x)} |_{x=x_0}\quad .
 \label{eq:exp}
\end{equation}
For the TFIMs investigated in this work one has $z=1$ for the dynamical critical exponent and the dlogPad\'e extrapolation of the one-particle gap directly yields the critical exponent $\nu$. Convergence of a physical quantity is suggested when different extrapolants $\left[L,M\right]$ give essentially the same result.


\begin{thebibliography}{30}
\bibitem [{Ext()}]{Extrapol}%
  \BibitemOpen
  \href@noop {} {}\bibinfo {note} {{We have determined the average value and
  its standard deviation of the critical field and of the critical exponent by
  taking into account dlogPad\'{e} extrapolants of the maximal perturbative
  order calculated \cite{Supp_Mat}.}}\BibitemShut {Stop}%
\end{thebibliography}
\end{document}